\documentclass[11pt, a4paper]{article}

\usepackage{jheppub}
\usepackage{amsmath}
\usepackage{amssymb}
\usepackage{graphicx}
\usepackage{slashed}
\usepackage{textcomp}

\usepackage{epstopdf}

\newcommand{\be}{\begin{eqnarray}}
\newcommand{\ee}{\end{eqnarray}}
\newcommand{\nn}{\nonumber}
\newcommand{\bn}{\begin{enumerate}}
\newcommand{\en}{\end{enumerate}}

\def\CD{{\cal D}}

\def\CG{{\cal G}}

\def\CL{{\cal L}}

\def\CN{{\cal N}}
\def\CO{{\cal O}}

\def\a{\alpha}

\def\g{\gamma}
\def\d{\delta}

\def\z{\zeta}

\def\k{\kappa}
\def\l{\lambda}
\def\m{\mu}
\def\n{\nu}

\def\r{\rho}

\def\s{\sigma}

\def\G{\Gamma}

\def\half{\frac{1}{2}}

\def\p{\partial}

\def\jmath{{j}}

\newcommand{\beq}{\begin{equation}}
\newcommand{\eeq}{\end{equation}}
\newcommand{\bea}{\begin{eqnarray}}
\newcommand{\eea}{\end{eqnarray}}
\newcommand{\ba}{\begin{array}}
\newcommand{\ea}{\end{array}}

\newcommand{\bg}{\bar{g}}

\newcommand{\blambda}{\bar{\lambda}}

\newcommand{\bpsi}{{\bar{\psi}}}

\newcommand{\vx}{\vec{x}}
\newcommand{\vy}{\vec{y}}

\title{Noncritical Einstein-Weyl Gravity and the AdS/CFT Correspondence}

\author[a]{Seungjoon Hyun,}
\author[a]{Wooje Jang,}
\author[a]{Jaehoon Jeong}
\author[b]{and Sang-Heon Yi}

\affiliation[a]{Department of Physics, College of Science, Yonsei University, Seoul 120-749, Korea}
\affiliation[b]{Center for Quantum Spacetime, Sogang University, Seoul 121-741, Korea}

\emailAdd{sjhyun@yonsei.ac.kr}
\emailAdd{wooje@yonsei.ac.kr}
\emailAdd{j.jeong@yonsei.ac.kr}
\emailAdd{shyi@sogang.ac.kr}

\abstract{ We explore four-dimensional Einstein-Weyl gravity and supergravity on anti-de Sitter spacetime. 
 For a specific range of the coupling with appropriate boundary conditions,  we show the effective equivalence of the theory with Einstein gravity and AdS supergravity at the quadratic Lagrangian level. Furthermore we show that these equivalences can be promoted to the full nonlinear level. We also show that the similar behavior holds for the generalized Gibbons-Hawking terms. From this we find that  the correlation functions in the dual conformal field theory of Einstein-Weyl gravity and supergravity can be readily read off from corresponding ones from Einstein gravity and AdS supergravity. We also give comments on some issues in critical gravity and supergravity as well as conformal gravity and supergravity.}
\arxivnumber{1111.1175}

\begin{document}
\maketitle

\section{Introduction}

Recently there have been much interests in higher derivative gravity. This renewed interests started with the following observations in the three-dimensional gravity. In three-dimensional Einstein gravity, there is no dynamical graviton which seems to make the theory rather trivial, though nontrivial black hole solutions exist.  However, the situations are changed when higher derivative terms are included. Specifically, propagating graviton modes appear when the, so-called, gravitational Chern-Simons term is added to the  three-dimensional pure Einstein gravity. Since the parity is broken due to the gravitational Chern-Simons term, the model contains the massive graviton with only one helicity. This theory  with negative cosmological constant  admits anti-de Sitter(AdS) spacetime  as a solution. Therefore one may study the dual two-dimensional conformal field theory(CFT) through the AdS/CFT correspondence\cite{Maldacena:1997re}\cite{Witten:1998qj}\cite{Gubser:1998bc}.  In particular, for  a specific value for the gravitational Chern-Simons coupling, the central charge of one chiral sector of the dual CFT vanishes. It was argued in
\cite{Li:2008dq}\cite{Maloney:2009ck} that the dual CFT should be  chiral, known as
chiral gravity conjecture. On the other hand,  it was claimed in \cite{Grumiller:2008qz} that the log modes exist in the bulk gravity  by relaxing the boundary condition and was conjectured that it is dual to the logarithmic CFT.

Another interesting three-dimensional gravity is new massive gravity realized in \cite{Bergshoeff:2009hq}. Its action contains a specific combination of $R^2$ and $R_{\m\n}R^{\m\n}$. In this theory  massive gravitons exist with both helicities, as the parity is conserved. Since it also admits AdS spacetime as a vacuum solution, new massive gravity on the AdS spacetime can be regarded as the dual of a certain two-dimensional CFT.  When the couplings of higher derivative terms take a special value, the dual CFT has  vanishing central charges in both sectors.

More recently, inspired by developments in three-dimensional new massive gravity,
a four-dimensional higher derivative gravity was proposed in \cite{Lu:2011zk}, which is called critical gravity. See also \cite{Deser:2011xc} $\sim$ \cite{Afshar:2011qw}. It is a higher derivative gravity whose action includes Einstein-Hilbert, cosmological constant  and Weyl-squared terms with a specific value of coupling. Generically, a four-dimensional theory with higher derivative terms contains massless and massive gravitons, one of which turns out be ghost-like. However critical gravity, due to the specific value of the coupling,  contains massless and logarithmic modes. The massless mode turn out to have zero excitation energy. 
It admits AdS Schwarzschild black hole solution whose mass and entropy vanish. 
These suggest that the inclusion of the logarithmic mode is needed to have nontrivial contents, but then the theory seems to be non-unitary and dual to the logarithmic CFT.

On the other hand it was argued in \cite{Maldacena:2011mk} that conformal gravity on AdS spacetime  may describe the same physics as pure Einstein gravity at low energy. Conformal gravity contains massless as well as massive gravitons. If this ghost-like massive mode falls off more slowly than massless mode, it may be truncated by the boundary condition, while leaving the massless graviton only.
Based on the study of the AdS Schwarzschild black hole it was claimed that the equivalence between conformal gravity and Einstein gravity holds even at the nonlinear level.

Subsequently in \cite{Lu:2011ks} they considered the higher derivative gravity on AdS spacetime with some specific range of the Weyl-squared coupling. This, so-called, Noncritical Einstein-Weyl(NEW) gravity includes the ghost-like massive mode which falls off more slowly than the massless mode and thus can be consistently truncated.   This was extended in \cite{Lu:2011mw} to $\CN =1$ NEW supergravity.

Though these studies are mostly based on the linearized equations of motion, they suggest that NEW gravity after consistent truncation may be well-defined at low energy. Furthermore they seem to indicate that NEW gravity may describe essentially the same physics as Einstein gravity.
In this paper we show the classical equivalence between NEW gravity and Einstein gravity on AdS background at the full nonlinear level. This is achieved by  showing that the effective Lagrangian of  NEW gravity becomes identical with Einstein gravity with the rescaling of Newton's constant. From this we find that  $n$-point correlation functions of energy-momentum tensors in the dual CFT can be read off from the results in Einstein gravity.
We also show the equivalence between ${\cal N}=1$ NEW supergravity and  ${\cal N}=1$ AdS supergravity, from which one can read off
 $n$-point correlation functions including energy-momentum tensors and  two supercurrents.

This paper is organized as follows. In section 2, we review critical gravity and NEW gravity.  We also present the auxiliary field formalism which makes the linearized level equivalence more transparent.  In section 3, we find the full effective action of NEW gravity at the tree level, which has the exactly same form as Einstein gravity with the rescaled Newton's constant. We show that the generalized Gibbons-Hawking boundary terms also has the same behavior as the bulk action.  From these we explain how to obtain  $n$-point correlation functions of energy-momentum tensors in the dual boundary CFT. In section 4, we extend our analysis to ${\cal N}=1$ NEW supergravity. In section 5, we give some comments on critical gravity and conformal gravity.
In section 6, we draw our conclusions.

\section{Critical Gravity and  NEW Gravity}

In this section we review critical gravity and NEW gravity.
Our starting point is the  higher curvature gravity which contains Einstein-Hilbert term with negative cosmological constant and Weyl-squared term.
The linearized field equation of metric becomes fourth order differential equation.  The natural vacuum solution is given by AdS spacetime.  If we expand the metric field on this AdS background,  we obtain two different modes, one  is the massless graviton and the other, generically,  ghost-like massive graviton or tachyon.

For the  special value of the coupling of Weyl-squared term, the theory is called critical gravity. In this theory, the linearized fluctuation modes consist of massless and logarithmic ones. The nonunitary logarithmic mode falls off more slowly than massless mode. This suggests that we may  consistently truncate the  logarithmic mode by choosing an appropriate boundary condition. Then, apparently, the theory has massless graviton only, just like pure Einstein gravity. The problem is that the excitation energy of this  graviton is zero, which seems to indicate that the content of the theory is null. It also admits the AdS Schwarzschild black hole solution whose mass and entropy turn out to be zero.  These seem to be partially explained by the observation\cite{Maldacena:2011mk} that conformal gravity on AdS spacetime, with critical value of the coupling, describes the same physics as Einstein gravity at low energy.

In NEW gravity,  the situation is changed. Once again we may take a consistent  truncation of the ghost-like massive graviton by choosing suitable boundary condition. In this theory, the massless graviton mode have the positive excitation energy and it becomes the theory of massless spin two field at low energy. Since the only nonlinear realization of massless spin two field is given by Einstein gravity, it seems to suggest that NEW gravity with consistent truncation of ghost-like massive mode is equivalent to Einstein gravity at low energy.

We begin with the  action which contains Weyl-squared terms as
\bea
 S&=&\frac{\s}{2\k^{2}}\int d^{4} x \sqrt{-g} \left[ R+\frac{6}{\ell^2} -\frac{1}{4m^2}C^{\m\n\r\s}C_{\m\n\r\s} \right] \\
 &=&\frac{\s}{2\k^{2}}\int d^{4} x \sqrt{-g} \left[ R+\frac{6}{\ell^2} -\frac{1}{2m^2}\Big( R^{\m\n}R_{\m\n} -\frac{1}{3} R^{2} \Big)-\frac{1}{4m^2} E\right] \, ,\nn
 \label{NEW}
\eea
where the Euler density, $E= R^{\m\n\r\s}R_{\m\n\r\s}-4R^{\m\n}R_{\m\n} +R^2$, is the total derivative in four dimensions. We choose $\ell^2>0$ where the theory admits AdS spacetime as a solution. On the other hand we allow $m^2$ to have an arbitrary real value and also the overall sign $\s=\pm 1$ of the total action yet to be determined.  In the usual Einstein gravity,  $\sigma$ should be positive, but with higher derivative terms, it may be allowed to take the negative value of $\s$.  Later on we will see that the choice, $\s m^2<0$, would be appropriate for our purpose.

The Euler-Lagrange equations of motion for the metric $g_{\m\n}$ are given by
\be
G_{\m\n}-\frac{3}{\ell^2} g_{\m\n}+E_{\m\n}=0~,
\ee
where
\bea
G_{\m\n}&=& R_{\m\n}-\frac{1}{2}R g_{\m\n} \, ,\\
E_{\m\n}&=& -\frac{1}{m^2} (R_{\m\r}R_{\n}^{~\r} - \frac{1}{4} R^{\r\s}R_{\r\s} g_{\m\n}) + \frac{1}{3m^2} R( R_{\m\n} - \frac{1}{4} R g_{\m\n}) \nn \\
&& - \frac{1}{2m^2} ( \Box R_{\m\n} + \half \Box R g_{\m\n} -2 \nabla_{\r} \nabla_{(\m} R_{\n)}^{~\r}) + \frac{1}{3m^2} ( g_{\m\n} \Box R - \nabla_{\m}\nabla_{\n} R)~,\nn
\eea
with $\Box\equiv \nabla^2$.

This theory admits AdS spacetime as the
vacuum solution of the field equations,
\be
 \bar R_{\m\n\r\s} = -\frac{1}{\ell^2}\left(\bar g_{\m\r} \bar g_{\n\s}-\bar g_{\m\s} \bar g_{\n\r}\right)\, , \quad \bar R_{\m\n} = -\frac{3}{\ell^2}\bar g_{\m\n} \, , \quad \bar R= -\frac{12}{\ell^2} \, .
 \label{AdS}
\ee
One may consider the fluctuation of the metric on this AdS spacetime as
\be
 g_{\m\n} = \bar{g}_{\m\n} +\d g_{\m\n} \, .
\ee
We can choose the
transverse traceless gauge, which is consistent with the equations of motion, as
\be
\bar \nabla^{\m} \d g_{\m\n} = 0 \,,\qquad \bg^{\m\n} \d g_{\m\n} = 0\,,
\ee
where $\bar\nabla_\m$ denotes the covariant derivative with respect to the background metric $\bg_{\m\n}$ and  the spacetime indices are raised/lowered by the background metric.
In this gauge, the linearized equation of motion becomes
\be
\left( \bar\Box + \frac{2}{\ell^2}\right) \left( \bar\Box + \frac{4}{\ell^2} - 2m^2 \right) \d g_{\m\n}=0\,.
\ee
Mode solutions of this fourth order differential equation consist of a massless graviton $h_{\m\n}$ which satisfy
\be
\left( \bar\Box + \frac{2}{\ell^2}\right)  h_{\m\n}=0\,,\label{eom1}
\ee
and a massive graviton $\phi_{\m\n}$, with mass $M$,
\be
 \left( \bar\Box + \frac{2}{\ell^2} - M^2 \right) \phi_{\m\n}=0\,, \quad M^2=2\Big(m^2-\frac{1}{\ell^2}\Big)\,.\label{eom2}
\ee
In the limit $m^2 = \frac{1}{\ell^2}$, the two quadratic differential operators degenerate. In this case, which is called  critical gravity, the second mode solution is not massive but logarithmic which satisfies the full quartic differential equation,
\be
\left( \bar\Box + \frac{2}{\ell^2}\right)^2  h^L_{\m\n}=0\,.
\ee
Since this logarithmic mode falls off more slowly than massless mode as it approaches the boundary, we may be able to truncate logarithmic mode by boundary condition, while leaving massless mode only. However in \cite{Lu:2011zk} it was noted that the excitations of this massless mode have zero energy. This seems to  indicate that the theory becomes trivial. (See also \cite{Liu:2009bk} for the $D=3$ case.)

More interesting case is the noncritical one with $  -\frac{1}{8\ell^2}\leq m^2 <
 \frac{1}{\ell^2}$ where the lower bound, $ m^2 = -\frac{1}{8\ell^2}$, comes from the tachyon-free condition, known as the BF bound\cite{Breitenlohner:1982jf}.  In this case, which we call NEW gravity, the theory is still tachyon-free, while containing ghost-like massive graviton. Furthermore since $M^2<0$ in the above coupling range, this massive mode falls off more slowly than massless mode as approaching the boundary and therefore  we can  truncate consistently the massive mode by taking appropriate boundary condition\cite{Maldacena:2011mk}\cite{Lu:2011ks}. The remaining massless graviton has non-zero excitation energy and thus describes gravity.

In order to deal with the higher derivative terms, it is convenient to introduce the auxiliary field.
We introduce an auxiliary field $f_{\m\n}$ with which the action can be written as the quadratic form,
\be
 S=\frac{\s}{2\k^{2}} \int d^{4} x \sqrt{-g} \left[ R + \frac{6}{\ell^2} -\half f^{\m\n} G_{\m\n} + \frac{m^{2}}{8} \left( f^{\m\n} f_{\m\n} - f^{2} \right) \right]~.
 \label{actionf}
\ee
The Euler-Lagrange equations for the auxiliary field $f_{\m\n}$ and the metric $g_{\m\n}$ are, respectively,  given by
\bea
f_{\mu\nu} &=& \frac{2}{m^2} \left( R_{\mu\nu} - \frac{1}{6} g_{\mu\nu} R \right)\,,\\
0 &=& G_{\mu\nu} - \frac{3}{\ell^2} g_{\mu\nu} -  f^{\rho}_{~(\mu}G_{\nu )\rho}
     - \frac{1}{4} \left[f_{\mu\nu} R - f^{\rho\sigma} R_{\rho \sigma} g_{\mu\nu} - f G_{\mu\nu}  \right]\nn\\
     && +\frac{1}{4} \left[ 2\nabla_{\rho}\nabla_{(\mu} f_{\nu)}^{~\rho}
         -  \nabla_{\rho}\nabla_{\sigma}f^{\rho\sigma} g_{\mu\nu} - \Box \left(f_{\mu\nu} - f g_{\mu\nu} \right)
         -  \nabla_{\mu}\nabla_{\nu}f \right]\nn\\
     && + \frac{m^2}{4} \left[ f^{\rho}_{~(\mu} f_{\nu)\rho} -  f f_{\mu\nu}
          - \frac{1}{4} \left(f^{\rho\sigma} f_{\rho\sigma} - f^2\right) g_{\mu\nu}\right]\,.\nn
\label{EOMA}
\eea

By plugging the expression of $f_{\m\n}$ in (\ref{EOMA}) back in the above action, we recover the original action (\ref{NEW}).
The vacuum solution of the equations of motion is given by 
AdS spacetime (\ref{AdS}) with
\be
 \bar f_{\m\n} = -\frac{2}{m^{2}\ell^2} \bg_{\m\n} \, , \quad \bar f = - \frac{8}{m^{2}\ell^2}  \, .
 \label{background}
\ee

\section{NEW Gravity and Its AdS/CFT  Correspondence}

In this section we explicitly construct the full nonlinear effective action of NEW gravity at the tree level. First of all, we show that  massless and massive modes are decoupled at the quadratic Lagrangian. Thus the ghost-like massive mode  can be consistently  truncated by an appropriate boundary condition. We show that the full nonlinear action, after the consistent truncation, becomes effectively the Einstein-Hilbert action with the cosmological constant. We also show that the generalized Gibbons-Hawking terms become effectively the one in Einstein gravity as well.  
This leads us to obtain the large $N$ limit of  $n$-point correlation functions of the energy-momentum tensors in the dual CFT via AdS/CFT correspondence.

\subsection{Effective quadratic Lagrangian}
We expand the metric $g_{\m\n}$ and the auxiliary field $f_{\m\n}$ around the background AdS metric $\bar{g}_{\m\n}$. A judicious choice of the expansion is given by
\be
 g_{\m\n} &=& \bar{g}_{\m\n} +  h_{\m\n} + \phi_{\m\n}\,, \\
 f_{\m\n} &=& \bar{f}_{\m\n} -\frac{2}{m^{2}\ell^{2}} h_{\m\n} -2 \phi_{\m\n} \nn\\
 &=& -\frac{2}{m^{2} \ell^{2}} \left( \bar{g}_{\m\n} + h_{\m\n} + m^{2} \ell^{2} \phi_{\m\n} \right)\,,\nn
\label{expand1}
\ee
which, as will be shown later on,  leads  to the natural decoupling of massless and massive modes.
The Ricci tensor can be expanded as
\beq
R_{\mu\nu} = \bar{R}_{\mu\nu} +  R_{\mu\nu}^{(1)}(h+\phi) +  R_{\mu\nu}^{(2)}(h+\phi)  + \cdots\,,
\eeq
where the first order term $R_{\mu\nu}^{(1)}$ is given by
\bea
R_{\mu\nu}^{(1)} (h)&=& \bar\nabla_\rho \bar\nabla_{(\mu} h^\rho_{\nu)} - \frac{1}{2}\bar\nabla_\nu \bar\nabla_\mu h - \frac{1}{2}\bar\Box h_{\mu\nu}\,.
\eea
The second order term $R_{\mu\nu}^{(2)}$, when contracted with the background metric, is given by
\bea
\bg^{\mu\nu}  R_{\mu\nu}^{(2)}(h) &=& \half h^{\mu\nu} \left( R_{\mu\nu}^{(1)}(h) - \half \bg_{\mu\nu}\bg^{\rho\sigma}  R_{\rho\sigma}^{(1)}(h)\right) + \bar\nabla_{\mu} \theta^{\mu}(h)\,,
\eea
where 
\bea
 \label{theta}
 \theta^{\mu}(h) = \frac{3}{4} h_{\rho\nu} \bar\nabla^{\mu} h^{\rho\nu}+\frac{3}{4} h^{\mu\nu}\bar\nabla_{\nu} h -\half h_{\rho\nu} \bar\nabla^{\rho} h^{\mu\nu} -h^{\mu\nu}\bar\nabla^{\rho}h_{\nu \rho}
+\frac{1}{4} h \bar\nabla_{\nu} h^{\mu\nu}  -\frac{1}{4} h \bar\nabla^{\mu} h.
\eea
 Accordingly  the action (\ref{actionf}) can be expanded and its quadratic part can be written as
\be
S_2=\frac{1}{2\k^{2}} \int d^{4} x \sqrt{-\bg}~ \CL_{2} \,,
\ee
 where the  quadratic  Lagrangian  $\CL_2$  is given by, modulo total derivatives,
\be
 \mathcal{L}_{2} = -\frac{q}{2}  h^{\m\n} {\cal{G}}_{\m\n}(h) +\frac{q}{2}   \phi^{\m\n} {\cal{G}}_{\m\n}(\phi)  + \frac{\s}{2} m^{2}q^{2} \left( \phi^{\m\n} \phi_{\m\n} - \phi^{2} \right)\,.
 \label{eff1}
\ee
Here ${\cal{G}}_{\m\n}$ denotes the linearized Einstein operator including the cosmological term,
\be
 {\cal{G}}_{\m\n}(h)
 = R^{(1)}_{\m\n}(h) -\half \bar{g}_{\m\n}\bar{g}^{\r\s} R^{(1)}_{\r\s}(h)  +\frac{3}{\ell^2} h_{\m\n} - \frac{3}{2\ell^2} h \bar{g}_{\m\n}\,,
\ee
 and a new parameter $q=\s\big(1-\frac{1}{m^{2}\ell^{2}}\big)$ is introduced.

The linearized equations of motion are given by
\be
  {\cal{G}}_{\m\n}(h) &=& 0\,,  \\
  {\cal{G}}_{\m\n}(\phi) &+& \s  m^{2} q\left( \phi_{\m\n} - \phi \bar{g}_{\m\n} \right) =0 \,,\nn
  \label{leom}
\ee
which correspond to (\ref{eom1}) and (\ref{eom2}), respectively.
From the above quadratic Lagrangian, it is clear that $ h_{\m\n}$ and $\phi_{\m\n}$
correspond to the massless and massive gravitons, respectively.  In these parametrization, the massless and massive modes are completely decoupled at the quadratic Lagrangian and therefore bulk-to-boundary propagators in the context of AdS/CFT correspondence do not mix those two kinds of modes.\footnote{ This is in contrast with the critical case, where there should be a mixing between the massless mode and logarithmic mode. See section 5.} Furthermore the signatures of their kinetic terms are opposite, signaling one of them is ghost-like mode. If we take $q$ positive, the massless graviton becomes the physical one, while the massive mode becomes ghost-like with wrong sign of kinetic term.  

The theory is tachyon-free if $ m^2\geq -\frac{1}{8\ell^2}  $, though it still  contains ghost-like massive mode.
But if  $ m^2 < \frac{1}{\ell^2}$, i.e. $M^2<0$, the ghost-like massive mode falls off more slowly than the massless graviton  and therefore can be consistently truncated by the boundary condition.
Henceforth we consider the parameter range $  -\frac{1}{8\ell^2}\leq m^2 < \frac{1}{\ell^2}$, and we have two cases\footnote{In \cite{Lu:2011ks}\cite{Lu:2011mw},  only case (I) is considered .} which maintain $q$ positive~:
\begin{itemize}
\item  Case (I) : $  -\frac{1}{8\ell^2}\leq m^2 \leq 0$ and $\sigma =1$.
\item Case (II) : $0 \leq m^2 < \frac{1}{\ell^2} $  and $\sigma =-1$.
\end{itemize}
One may note that $\s m^2$, which is proportional to the coupling of the Weyl-squared term, should be always negative. Critical gravity corresponds to the case $q=0$ or $m^2\ell^2=1$, which we need to deal with separately. Conformal gravity corresponds to the limit $m\rightarrow 0_\pm$.

 In the context of AdS/CFT correspondence, this truncation of  the massive mode by the boundary condition corresponds to turning off the source for the operator dual to the ghost-like massive mode. Our quadratic Lagrangian tells us that once the massive mode is removed at the boundary, that would persist deep in the bulk.


\subsection{Effective boundary action}
In this section we describe the effective boundary action and find the similar behavior.
 As stressed in the previous section,
 since the ghost-like massive mode can be consistently truncated,   we restrict ourselves to the case when $\phi_{\m\n}=0$ from now on.

We have two different sources for the boundary action $S_{bndry}$, one coming from the
on-shell bulk action and the other from the generalized Gibbons-Hawking terms,
 \be
 S_{bndry}= S|_{on-shell}+S_{GGH}\,.
 \ee
The on-shell bulk action is given by
\be
 S|_{on-shell} = \frac{1}{2\k^{2}} \int_{M} d^{4} x \sqrt{\bg} \left(\bar\nabla_{\m} \xi^{\m}+\CO (h^3)  \right)\,,
\ee
where
\be
 \xi^{\m} = q \left[ \bar\nabla_{\n}  h^{\m\n}  -\bar\nabla^{\m}  h \right] + q \theta^{\m}(h) \,,
\ee
with $\theta^\mu$ in (\ref{theta}).

The generalized Gibbons-Hawking terms are determined by the requirement of the well-defined variational principle and can be shown to be\cite{Hohm:2010jc}
\bea
 S_{GGH} &=& \frac{\s}{2\k^{2}} \int_{\partial M} d^{3} x \sqrt{\g} \left(-2 K +\half \hat{f}^{ij}K_{ij} -\half \hat{f} K \right)\,,
\eea
where $\gamma_{ij}$ denotes the metric at the boundary and $\hat{f}_{ij}$ and $\hat{f}=\hat{f}_{ij}\gamma^{ij}$ are defined in \cite{Hohm:2010jc}.
Note that
\beq K_{ij} = -  \nabla_{(i} n_{j)} \,, \qquad K = \g^{ij} K_{ij}\,, \eeq
where $n_{\m}$ is the unit normal vector at the boundary.
These boundary actions can be used to find the correlation functions of the boundary energy-momentum tensor.

To be concrete we choose background Euclidean AdS spacetime in Poincar\'e coordinates,
\be
 ds^{2} = \bg_{\mu \nu}dx^\mu dx^\nu=\frac{\ell^{2}}{x_{0}^{2}} \left( dx_{0}^{2} +dx_{1}^{2} +dx_{2}^{2} +dx_{3}^{2}  \right) .  \label{EAdS}
\ee
  In these coordinates, it turns out that $\hat{f}_{ij}=f_{ij}$.
 An appropriate gauge condition at the boundary is chosen as
\be
 h_{00} =h_{0i} =0 \,,
\ee
which holds in the bulk as well\cite{Liu:1998bu}.
The expansion of the boundary metric is given by 
\be
\g_{ij} = \frac{\ell^{2}}{x_{0}^{2}}(\delta_{ij} + h^B_{ij})\,,
\ee
where $\delta_{ij}$ denotes the metric of the flat CFT background and $h^B_{ij}$ does the metric fluctuation. 
Since  $\hat{f}_{ij}=f_{ij}$ in Poincar\'e coordinates, the boundary fields $\hat{f}_{ij}$ and $\hat{f}$ are expanded in the same way as  the  bulk auxiliary field $f_{\m\n}$ given in (\ref{expand1}),
\bea
\hat{f}_{ij} = -\frac{2}{m^2 x_{0}^{2}}(\delta_{ij} + h^B_{ij})\,,\qquad
\hat{f} = -\frac{6}{m^2\ell^2}\,.
\label{expand2}
\eea
The on-shell Euclidean boundary action, $I_E= - S_{bndry}$, becomes
\beq
 I_{E} =\frac{q \ell^2}{2\k^{2}} \int_{\partial M} d^{3} x x_0^{-2} \left[ \frac{2}{x_0} \hat{h} + \left( \frac{1}{2x_0}\hat{h}^2 + \frac{1}{4} h^{B}_{ij} \partial_{0} h^{B}_{ij} - \frac{1}{4}\hat{h} \partial_{0} \hat{h} \right)+\CO (h^{3}_{B}) \right]\,,
 \label{bndry action}
 \eeq
where $\hat{h} = \delta^{ij} h^B_{ij}$.
One may note that the action is exactly the same as the one from Einstein gravity on AdS spacetime with the rescaling of $\kappa^2$ by $q$.

\subsection{CFT correlation functions and nonlinear completion in NEW gravity}

In this section we find the bulk-to-boundary propagator and find the two-point correlation function of the energy-momentum tensor.
Then we show that the equivalence between NEW gravity and Einstein gravity at the quadratic level, in the previous section, can be promoted to full nonlinear level.  We also show that the quadratic level equivalence in the boundary action can be promoted to full nonlinear level as well. This equivalence implies  the equivalence of $n$-point correlation functions in the dual CFTs.

The bulk field $h_{ij}(x^0,\vec{x})$ is related to the boundary source $h^B_{ij}$ of the energy-momentum tensor by the bulk-to-boundary propagator as\cite{Liu:1998bu}
\beq
h_{ij} (x_0, \vec{x}) = \frac{8}{\pi^2 } \int d^3 y \frac{x_0^3}{(x_0^2+|\vx - \vy|^2)^3}
                               J_{ik}(x_0,\vec{x} - \vec{y}) J_{jl}(x_0,\vec{x} - \vec{y}) P_{klmn} \hat{h}_{mn}(\vy)\,,
\eeq
where
\beq
P_{ijkl} = \half (\d_{ik} \d_{jl} + \d_{jk} \d_{il}) - \frac{1}{3} \d_{ij} \d_{kl}\,, \qquad
 J_{ij} ( x_0, \vec{x} ) = \d_{ij} - \frac{2 x_i x_j}{x_0^2+|\vec{x}|^2}\,.\qquad
\eeq
After the appropriate boundary local counter terms, for the cancellation of the divergence, are included, the remaining
quadratic action becomes
\bea
I_E
   &=& \frac{3 q \ell^2}{\pi^2 \kappa^2} \int_{\partial M} d^{3} x d^3 y
       \frac{\hat{h}_{ij}(\vec{x}) H_{ijkl} (\vx - \vy) \hat{h}_{kl}(\vec{y})}{|\vec{x} - \vec{y} |^6}\,,
\eea
where
\beq
H_{ijkl}(\vx) \equiv \half(J_{ik}J_{jl} + J_{il} J_{jk} )\big|_{x_0=0} - \frac{1}{3} \d_{ij} \d_{kl}\,.
\eeq
Therefore the two point correlation function of the energy-momentum is given by
\beq
\Big\langle T_{ij}(\vx) T_{kl}(\vy)\Big\rangle = C_T \frac{ H_{ijkl} (\vx - \vy) }{|\vx- \vy |^6}~,
\eeq
with the central charge of the dual CFT,
\beq
C_T = \frac{3 q \ell^2}{\pi^2 \kappa^2}\,.
\eeq
Note that in the critical limit $q\rightarrow0$, the central charge vanishes, signaling the null content of the dual CFT. In order to have the nontrivial contents in the critical limit it seems to be essential that we include the log mode.

Once the ghost-like massive mode is truncated, our discussion at the linearized level can be promoted to full nonlinear level. Without massive mode, we may promote the linearized relation (\ref{expand1}) to the full nonlinear level as
\beq
f_{\m\n} = - \frac{2}{m^2 \ell^2} g_{\m\n} \,,  \quad f = - \frac{8}{m^{2}\ell^2}  \,,
\label{nonlinear1}
\eeq
where $g_{\m\n}$ contains only massless mode.
If we plug this relation back into the original action of NEW gravity, we obtain
\be
 S=\frac{q}{2\k^{2}} \int d^{4} x \sqrt{-g} \left[ R + \frac{6}{\ell^2} \right]~,
\ee
which is nothing but the Einstein-Hilbert action with the same cosmological constant!  The only difference is the rescaling of the Newton's constant by $q$. Moreover the expansion (\ref{expand2}) of the  boundary field $\hat{f}_{ij}$ can be promoted to the full nonlinear level as
\bea
\hat{f}_{ij} = -\frac{2}{m^2\ell^2} \g_{ij} \,,\qquad
\hat{f} = -\frac{6}{m^2\ell^2}\,.
\eea
Then the generalized Gibbons-Hawking terms turn into the ordinary Gibbons-Hawking term with rescaled Newton's constant as
\bea
 S_{GGH} &=& -\frac{q}{\k^{2}} \int_{\partial M} d^{3} x \sqrt{\g}\,  K \,.
\eea

Via the AdS/CFT correspondence, the large $N$ limit in the dual CFT corresponds to taking the on-shell value of the  classical action in bulk gravity. This tells us that, in the large $N$ limit,  general $n$-point correlation functions of energy-momentum tensor in the dual CFT have the exactly same form as those from Einstein gravity on AdS spacetime with  the overall rescaling  by a factor of $q$.

\begin{figure}[htbp]
\begin{center}
\includegraphics[scale=0.3]{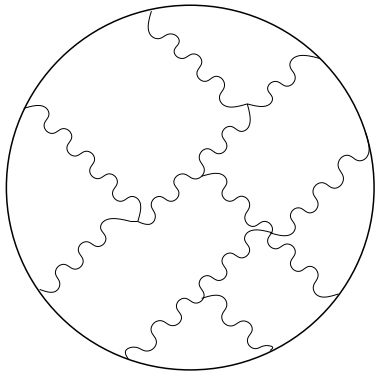}
\caption{Tree level Witten diagram}
\label{default}
\end{center}
\end{figure}


\section{ $\CN=1$ NEW Supergravity}
In this section  we consider ${\cal N}=1$ NEW supergravity.   In this supersymmetric extension of NEW gravity, we again find the similar behaviors as in NEW gravity. Generically it contains the massless  and the massive supermultiplet. We obtain the quadratic level effective action in which these modes are decoupled. The dynamical fields in this  theory consist of the metric, massive vector  and gravitino fields. The metric has two modes, one massless and the other ghost-like massive modes. On the other hand, the gravitino field contains one massless mode and two massive modes. Those ghost-like massive graviton and gravitino can be consistently truncated by the appropriate boundary conditions leaving the physical massless multiplet and, if any, the well-behaved physical massive multiplet.
\subsection{Effective quadratic Lagrangian}

We begin with the ${\cal N}=1$ NEW supergravity action,
\begin{equation}
S= \frac{1}{2\k^{2}} \int d^{4} x \sqrt{-g} \left[\CL_B + \CL_F\right]~.
\end{equation}
The bosonic  Lagrangian $\CL_B$ is given by
\bea
 \CL_B &=& \s \left( \CL_{B}^{1} + \CL_{B}^{2} \right)\,,\\
 \CL_{B}^{1} &=& R + \frac{6}{\ell^2} +\frac{2}{3}A_\m A^\m \,,\nn \\
 \CL_{B}^{2} &=&-\half f^{\m\n} G_{\m\n} + \frac{m^{2}}{8} \left( f^{\m\n} f_{\m\n} - f^{2} \right)+\frac{1}{6m^2}F_{\m\n}F^{\m\n}~.\nn
\eea
The Lagrangian $ \CL_{B}^{1}$  is one for bosonic fields in the  Einstein and the cosmological constant supermultiplet,
 while the Lagrangian $ \CL_{B}^{2}$ is one for the bosonic fields in  the Weyl supermultiplet.
One may note that we plugged the on-shell value of all the auxiliary fields except for $f_{\m\n}$ in the original ${\cal N}=1$ off-shell supergravity action. The auxiliary $f_{\m\n}$ is the same as the one in \cite{Fradkin:1985am} when we restrict to the symmetric bosonic part only.  

By including Weyl-squared term the auxiliary vector field $A_\m$ becomes dynamical and turns into massive vector field. In our choice of the parametrization, the kinetic term for the vector field $A_\m$ has always the right sign, while its mass term can have  both signs. In the case (I),  $m^2$ is negative and therefore the vector field is ghost-like. This ghost-like vector field, as it falls off more slowly than the massless fields, can be consistently truncated by the boundary condition.  In the case (II), $m^2$ is positive and thus the vector field is physical and massive, which falls off more rapidly than the massless fields as it approaches to the boundary.
In this case it can not be truncated by the boundary condition. 

The fermionic Lagrangian $\CL_F$ is given by, up to quadratic order in the fermionic fields,
\bea
\CL_F &=& \s \left( \CL_{F}^{1} + \CL_{F}^{2} \right)\,,\\
\CL_{F}^{1} &=& 2\bpsi_{\m} \g^{\m\n\r} D_{\n} \psi_{\r} - \frac{2}{\ell} \bpsi_{\m} \g^{\m\n} \psi_{\n} +{\cal O}(\psi^4)\,,\nn\\
\CL_{F}^{2} &=& - \frac{4}{m^2} \bar{\chi}_{\mu}\gamma^{\mu\nu\rho}D_{\nu}\chi_{\rho} + \frac{2}{m^2} \bar{\z}^{\m}
                 ( \chi_{\m} - \chi_{0 \m})   - \half (\nabla_{\rho}f_{ \mu\nu})\bar{\psi}^{\mu}\gamma^{\nu}\psi^{\rho}\nn\\
&& - \frac{3}{2} \bar{\psi}^{\mu}f^{ \rho}_{~[\mu}\gamma^{}_{\rho\lambda]}\chi^{\lambda} - f_{ \mu\nu}\bar{\psi}_{\lambda}\gamma^{\nu}(D^{[\mu}\psi^{\lambda]} -\gamma^{[\mu}\chi^{\lambda]}_0)  + {\cal O}(\psi^4)\,,\nn
\eea
where $\chi_0$ is defined as
\beq
\chi_{0\,\m} \equiv \frac{2}{3}\Big( \g^{\n} D_{[\nu}\psi_{\mu]}  + \frac{1}{4} \g_{\m\n\r} D^{[\nu}\psi^{\rho]}  \Big)\,.
\eeq
The Rarita-Schwinger Lagrangian $ \CL_{F}^{1}$  is one for fermionic fields in the  Einstein  and the cosmological constant supermultiplet,
 while the Lagrangian $ \CL_{F}^{2}$ is one for the fermionic fields in  the Weyl supermultiplet.
We introduced  auxiliary fermionic fields $\chi_\m$ and $\zeta_\m$ which would be considered as the superpartner of the bosonic auxiliary field $f_{\m\n}$.

The Euler-Lagrange equations for the bosonic fields are
\bea
G_{\m\n}-\frac{3}{\ell^2}g_{\m\n}&=&{\cal T}_{\m\n} (g, f ,A, \psi, \chi)\,,\\
f_{\mu\nu} &=& \frac{2}{m^2} \left( R_{\mu\nu} - \frac{1}{6} g_{\mu\nu} R \right)+\Delta_{\mu\nu}(g, A, \psi, \chi)~, \nn \\
\nabla_\m F^{\m\n}&=&2m^2 A^\n \,,\nn
\label{EOM1}
\eea
where ${\cal T}_{\m\n} (g, f ,A, \psi, \chi)$ contains purely bosonic terms shown in (\ref{EOMA}) and the energy-momentum tensor for $A_\m$ and quadratic and higher order fermionic terms while $\Delta_{\mu\nu}(g, A, \psi, \chi)$ consists of quadratic and higher order fermionic terms only. One may note that the auxiliary field $f_{\m\n}$ introduced in our paper is different from the one in \cite{Fradkin:1985am} by quadratic and higher order fermionic terms.

The Euler-Lagrange equations for the fermionic fields are given by
\bea
\chi_{\m} &=& \chi_{0 \m} \equiv \frac{2}{3}\Big( \g^{\n} D_{[\nu}\psi_{\mu]}  + \frac{1}{4} \g_{\m\n\r} D^{[\nu}\psi^{\rho]}  \Big) \,,\\
 \z_{\m} &=& 4\g_{\m\n\r} D^{\n} \chi^{\r} + \frac{3m^2}{4} f^{\n}_{~[\r} \g_{\m \n]}\psi^{\r} \,,\nn\\
0 &=& 4 ( \g_{\m\n\r} D^{\n} \psi^{\r} - \frac{1}{\ell} \g_{\m\n}\psi^{\n})
              + \frac{4}{3m^2} \g_{[\m} D^{\n} \z_{\n]}  - \frac{1}{3m^2} \g_{\m\n\r} D^{\n} \z^{\r}
           + \frac{3}{2}f^{\r}_{~[\m} \g^{}_{\n \r]}\chi^{\n}\nn\\
      &&  + \nabla_{[\m}f_{\n]\r} \g^{\r} \psi^{\n}
            + f^{\n\r}\g_{\r} ( D_{[\m} \psi_{\n]} - \g_{[\m} \chi_{0\n]}) \nn\\
         &&    + \frac{1}{6} D^{\n} \left[ 3f_{\r[\m}\g^{\r} \psi_{\n]}
            -f_{\r[\m}\g_{\n]}\psi^{\r} + 3f_{\r[\m}\g^{~~\r\s }_{\n]} \psi_{\s}
            -  f^{\r\s} \g_{\m\n\r} \psi_{\s} -  f \g_{[\m} \psi_{\n]} +  f \g_{\m\n\r} \psi^{\r} \right] \,. \nn
\eea
After plugging the equations of motion of the auxiliary fields $f_{\m\n}$, $\chi_\m$ and $\zeta_\m$
back into the action, we recover the original action of ${\cal N}=1$ Einstein-Weyl supergravity given in \cite{Fradkin:1985am} (See also \cite{Lu:2011mw}).

We expand the metric $g_{\m\n}$ as (\ref{expand1}) and obtain the same result (\ref{eff1}) as in NEW gravity. Now we expand the gravitino field $\psi_\m$   in the similar fashion.  A suitable choice which guarantees the decoupling among the massless mode $\psi^{(0)}_{\m}$ and massive modes $\psi^{(\pm)}_{\m}$ is 
\bea
\psi_{\m} &=& \psi^{(0)}_{\m} + \psi^{(+)}_{\m} + \psi^{(-)}_{\m}\,,\nn\\
\chi_{\m} &=& - \frac{1}{2\ell} \psi^{(0)}_{\m} -\frac{1}{2L_+} \psi^{(+)}_{\m} -\frac{1}{2L_-} \psi^{(-)}_{\m}\,,\\
\z^{\m} &=& - \frac{1}{\ell^2} \g^{\m\n}\psi_{\n}^{(0)} + \left( \frac{1}{\ell^2} - \frac{2}{L_+^2}\right)\g^{\m\n}\psi_{\n}^{(+)} +  \left( \frac{1}{\ell^2} - \frac{2}{L_-^2}\right)\g^{\m\n}\psi_{\n}^{(-)}\,, \nn
\eea
where
\beq
 \frac{\ell }{L_\pm} = \frac{1}{2} \left( -1 \pm \sqrt{1 + 8 m^2\ell^2} \right)\,.
\eeq

The quadratic  Lagrangian for fermion fields in the AdS background along with (\ref{background}) becomes
\bea
\CL_{F} = 2 q\, \bpsi_{\m}^{(0)} \bar\CD^{\m\n} \psi_{\n}^{(0)} +2 q_+ \bar\psi^{(+)}_{\m} \bar\CD_{+}^{\m\n} \psi_{\n}^{(+)}
    +2 q_- \bar\psi^{(-)}_{\m}\bar\CD_{-}^{\m\n} \psi_{\n}^{(-)} \,,
\eea
where
\be
\bar\CD^{\m\n}&=&\g^{\m\r\n}\bar D_{\r}  - \frac{1}{\ell}\g^{\m\n}\,,\quad\bar\CD_{\pm}^{\m\n}=\g^{\m\r\n}\bar D_{\r}  - \frac{1}{L_{\pm}}\g^{\m\n}\,,
\ee
and
\be
 q_\pm &=& \s \left( 1 + \frac{1}{2m^2\ell^2} - \frac{3}{2m^2 L_\pm^2} \right)\,.
\ee
The Euler-Lagrange equations for the massless mode $ \psi_{\n}^{(0)}$ and the massive ones $\psi_{\n}^{(\pm)}$ with masses $
M_{(\pm)}^2= \frac{1}{L_\pm^2}-\frac{1}{\ell^2}
$ are determined to be
\beq
\bar\CD^{\m\n} \psi_{\n}^{(0)} =0 \,,\qquad \bar\CD_{\pm}^{\m\n} \psi_{\n}^{(\pm)}=0\,.
\eeq

Just like in the bosonic case, these massless and massive modes are  completely decoupled in the quadratic Lagrangian and therefore the bulk-to-boundary propagators  do not mix among those modes. For the parameter range $ -\frac{1}{8\ell^2} < m^2 < \frac{1}{\ell^2} $ with the corresponding sign for $\s$ given in section 3.1 the coefficient $q_+$ becomes negative, while  $q_-$ remains positive. This tells us that the massive mode  $\psi^{(+)}$ is ghost-like.  For the $\psi^{(-)}$ we need to consider cases (I) and (II) separately. In the case (I), except for the special value $m^2= -\frac{1}{8\ell^2}$,  it becomes ghost-like, while in the case (II) it is massive and physical. In the case when  $m^2= -\frac{1}{8\ell^2}$, both coefficients, $q_\pm$, become zero as well as  the masses of those two massive gravitinos degenerate  indicating   one of  which should be the massive log mode.

\subsection{$\CN=1$ NEW supergravity and its AdS/CFT correspondence}
$\CN=1$ NEW supergravity with the case (I) contains the massless  supermultiplet $(h_{\m\n}, \psi_{\m}^{(0)})$
and the ghost-like massive supermultiplet $(\phi_{\m\n}, \psi_\m^{(\pm)}, A_{\m})$. Once again these ghost-like massive modes fall off more slowly than the massless modes, and  can be consistently truncated  by boundary conditions.
In AdS/CFT correspondence, this truncation corresponds to turning off the sources for the operators dual to the ghost-like  massive modes. Our quadratic Lagrangian tells us that once these ghost-like massive modes are removed at the boundary, that the truncation would persist in the bulk. From the above quadratic Lagrangian along with the appropriate boundary action, we can readily read off the two-point correlation function of supercurrent at the boundary. Our results strongly suggest that it is nothing but the supercurrent, $\xi_j$, two-point correlation function \cite{Corley:1998qg}\cite{Volovich:1998tj}\cite{Rashkov:1999ji} in $\CN=1$ AdS supergravity rescaled by the factor $q$:   
\beq
\Big\langle\xi_{i} (\vec{x}) \bar\xi_{j} (\vec{y}) \Big\rangle = \frac{4 q}{\pi^2 \k^2} \left( \d_{ik} - \frac{1}{3} \g_{i}\g_{k} \right) \frac{\vec{\g} \cdot ( \vec{x} - \vec{y})}{|\vec{x} - \vec{y} |^{6}} \left[ \d_{kj} - 2\frac{(x_k - y_k) (x_j - y_j)}{|\vec{x} - \vec{y} |^2}\right]\,.
\eeq

On the other hand, $\CN=1$ NEW supergravity with the case (II) contains the massless  supermultiplet $(h_{\m\n}, \psi_{\m}^{(0)})$
along with the ghost-like massive modes $(\phi_{\m\n}, \psi_{\m}^{(+)})$ and the physical massive modes $(A_{\m}, \psi_{\m}^{(-)})$. Since the physical massive modes fall off more rapidly than the massless modes as approaching the boundary, these can not be truncated. Therefore   $\CN=1$ NEW supergravity with the case (II) after the truncation includes
massless modes $(h_{\m\n}, \psi_{\m}^{(0)})$ and massive modes $(A_{\m}, \psi_{\m}^{(-)})$.

We showed that at the quadratic order
$\CN=1$ NEW supergravity with the case (I) after the truncation is equivalent to the usual $\CN=1$ AdS supergravity  modulo the rescaled Newton's constant by the factor $q$.
This equivalence can be promoted to the full nonlinear level in the metric and the quadratic level for gravitino by the nonlinear completion of the bosonic  fields $g_{\m\n}$ and  $f_{\m\n}$ as
\beq
f_{\m\n} = - \frac{2}{m^2 \ell^2} g_{\m\n} \,,  \quad f = - \frac{8}{m^{2}\ell^2}  \,,
\eeq
and of the fermionic fields as 
\bea
\chi_{\m} = - \frac{1}{2\ell} \psi_{\m} \,,\quad
\z^{\m} = - \frac{1}{\ell^2} \g^{\m\n}\psi_{\n} \,.
\eea
where $g_{\m\n}$ and $\psi_{\m} $ include massless mode only.
If we use this completion into the original action of $\CN=1$ NEW supergravity, we obtain
\be
 S=\frac{q}{2\k^{2}} \int d^{4} x \sqrt{-g} \left[ R + \frac{6}{\ell^2} +2\bpsi_{\m} \g^{\m\n\r} D_{\n} \psi_{\r} - \frac{2}{\ell} \bpsi_{\m} \g^{\m\n} \psi_{\n} +{\cal O}(\psi^4)\right]~.
\ee
Because of the supersymmetry, the quartic and higher order fermion terms may be guaranteed  to be the same as  $\CN=1$ AdS supergravity. This tells us that in the large $N$ limit the general $n$-point correlation functions among energy-momentum tensor and/or supercurrent in the dual CFT should be the same as those from  $\CN=1$ AdS supergravity with the rescaling by $q$.

\section{Comments on Critical Gravity and Conformal Gravity }

Critical gravity and $\CN=1$ critical supergravity correspond to the case when  $m^2\ell^2 = 1$ or $q=0$.  In this theory the appropriate expansion of  the metric $g_{\m\n}$ and $f_{\m\n}$ around AdS background is\cite{Bergshoeff:2011ri}
\bea
 g_{\m\n}=\bar{g}_{\m\n} + h_{\m\n}^L\,,\qquad
 f_{\m\n} = -\frac{2}{m^{2}\ell^{2}}(\bar{g}_{\m\n} +h^L_{\m\n}) + h_{\m\n}\,,
\eea
and then the quadratic part of the effective  bosonic Lagrangian  becomes
\beq
 \CL_{B} = \s \Big( \frac{1}{2} h^{\m\n}\CG_{\m\n} (h^L) -\frac{m^{2}}{4}\left( h^{\m\n} h_{\m\n} -h^{2} \right) + \frac{1}{6m^2} F_{\m\n} F^{\m\n} + \frac{2}{3} A_{\m} A^{\m}\Big)\,.
\eeq
The Euler-Lagrange equations at the critical point $q=0$ are given by
\bea
 \CG_{\m\n} (h)&=&0\,,\quad
 \CG_{\m\n} (h^L) = m^{2} (h_{\m\n} -h \bar{g}_{\m\n})\,,\quad
 \bar{\nabla}_{\m} F^{\m\n} = 2m^2 A^{\n}\,.
\eea
If we choose $\s=1$ then the bosonic part of critical supergravity includes massless and logarithmic gravitons and ghost-like massive vector field. It was noted that the excitations of massless modes  have zero energy indicating the null content of the theory after the truncation of the log mode. This also seems to be clear from the above expansion. If we set $h_{\m\n}^L=0$, the metric is fixed with the background value $g_{\m\n}=\bar{g}_{\m\n}$ while the auxiliary field $f_{\m\n}=-\frac{2}{m^{2}\ell^{2}}\bar{g}_{\m\n}+ h_{\m\n}$. If we plug these back in the action, there is no dynamical field  indicating the null content of the theory.  Therefore, after the truncation of the log mode and the ghost-like massive vector field, $\CN=1$ critical supergravity wouldn't have any content in the bosonic part. Shortly, it will be shown that it doesn't have any fermionic content either.  On the other hand if we choose $\s=-1$, the massive vector field becomes physical.  In this case after the truncation of the log mode,  the theory still has a massive vector field.

The fermionic fields can be expanded similarly as 
\bea
 \psi_{\m} &=& \psi_{\m}^{L} + \psi _{\m}^{-}\,,\nn\\
 \chi_{\m} &=& -\frac{1}{2\ell}\psi_{\m}^{L} + \frac{1}{\ell} \psi_{\m}^{0}+ \frac{1}{\ell}\psi_{\m}^{-} \,, \\
 \zeta_{\m} &=&-\frac{1}{\ell^{2}} \g^{\m\n}\psi_{\n}^{L} + \frac{8}{\ell^{2}}\g^{\m\n} \psi_{\n}^{0} -\frac{7}{\ell^{2}} \g^{\m\n} \psi_{\m}^{-}~,\nn
\eea
so that the massless mode $\psi_\m^{0}$ and associated log mode $ \psi_{\m}^{L} $ can be decoupled with the massive mode $\psi_\m^{-}$.
The effective quadratic  
 Lagrangian becomes
\beq
\CL_{F}=\s \Big(12 \bpsi^{0}_{\m} \bar{\CD}^{\m\n}\psi^{L}_{\n} - 4 \bpsi^{0}_{\m} \bar{\CD}^{\m\n}\psi^{0}_{\n} - 9 \bar\psi^{-}_{\m}\bar\CD_{-}^{\m\n} \psi_{\n}^{-} + \frac{12}{\ell} \bar{\psi}_{\m}^{0} \g^{\m\n} \psi_{\n}^{0} \Big)~,
\eeq
from which the Euler-Lagrange equations can be determined as
\bea
 \bar{\CD}^{\m\n} \psi^{0}_{\n} = 0\,,\quad
 \bar{\CD}^{\m\n}\psi^{L}_{\n} = - \frac{2}{\ell} \g^{\m\n} \psi^{0}_{\n}\,,\quad
 \bar\CD_{-}^{\m\n} \psi_{\n}^{-} =0\,.
\eea
These fermionic part shares common properties with the bosonic part. We expect that the massless mode has the vanishing excitation energy from the supersymmetry.  When $\s=1$, the massive mode $ \psi_{\m}^{-} $ has wrong sign in the kinetic term and thus becomes ghost-like. Therefore the theory might be null after the truncation of the log mode 
and the ghost-like massive mode. On the other hand, for  $\s=-1$, the massive mode $ \psi_{\m}^{-} $  becomes physical. Therefore in this case  $\CN=1$ critical supergravity after the truncation of the log modes may contain massive fields ($ A_\m, \psi_{\m}^{-} $).

Conformal gravity and conformal supergravity on AdS background correspond to the case when  we take the limit $m^2\rightarrow 0$, while the conformal coupling constant $\alpha^2=-4\s\k^2 m^2$ is kept fixed. The effective Lagrangians for NEW gravity and $\CN=1$ NEW supergravity in previous sections have the smooth limit.  Conformal gravity, after the truncation of ghost-like massive modes by boundary conditions,  contains massless graviton only and satisfies the same quadratic Lagrangian as Einstein gravity, with the replacement of Newton's constant $\k^2$ by conformal coupling $\a^2$. We can adopt the same method as was done in NEW gravity, namely the completion given in (\ref{nonlinear1}),  to promote the quadratic level  equivalence to the full level one. Therefore the effective action becomes 
\be
 S=\frac{2}{\a^{2}} \int d^{4} x \sqrt{-g} \left[ R + \frac{6}{\ell^2} \right]\,,
\ee
 which is  the pure Einstein gravity with cosmological constant. The same conclusion was drawn in \cite{Maldacena:2011mk}.\footnote{After the submission of this paper, we have received an e-mail from R. Metsaev informing that  he got the same result. (See \cite{Metsaev}.)}
In the conformal limit of $\CN=1$ NEW supergravity,  the parameters in the fermionic Lagrangian goes as $L_+\rightarrow \infty$ and $ L_- \rightarrow -\ell$.   Therefore $\CN=1$ conformal supergravity on AdS spacetime contains the massless multiplet consisting of one graviton, one gauge field  and two gravitinos. It also contains ghost-like massive graviton and gravitino. After the truncation of ghost-like massive modes by the boundary condition, the theory seems to be described by $\CN=2$ AdS supergravity, as one can see from the counting of the degrees of freedom and the our effective Lagrangian. 
In the absence of the background gauge field we may still use (\ref{nonlinear1}) to promote the equivalence to the full nonlinear level.

\section{Conclusions}
We studied four-dimensional Noncritical Einstein-Weyl(NEW) gravity and $\CN=1$ NEW supergravity, which are Einstein-Weyl gravity and supergravity on AdS spacetime with a specific range of the coupling constant. In this theory the ghost-like massive fields can be consistently truncated by  appropriate boundary conditions.  
By studying the effective Lagrangian, we showed 
that  the massless modes and ghost-like massive modes are decoupled  at the quadratic level and thus these truncations can be made consistently in the bulk. Then the effective quadratic Lagrangians of  NEW gravity and supergravity after the truncations become identical to those of Einstein gravity and AdS supergravity. We also show that this equivalence can be promoted to the full nonlinear level. Moreover this equivalence holds on the boundary action in which the generalized Gibbons-Hawking terms effectively reduce to the ordinary Gibbons-Hawking term.

This equivalence on the bulk gravity side guarantees that correlation functions in the dual CFT of NEW gravity and $\CN=1$ NEW supergravity can be readily read off from corresponding ones from Einstein gravity and $\CN=1$ AdS supergravity.
One may study the boundary CFT dual to NEW gravity and supergravity directly with higher curvature action, without introducing the auxiliary fields. We expect that correlation functions of the energy-momentum tensor multiplet related with the massless modes  will be the same as those from Einstein gravity and AdS supergravity.

It would be interesting to study quantum aspects on NEW gravity and $\CN=1$ NEW supergravity and how far our classical equivalence persists at the quantum level. We anticipate the equivalence holding at one-loop in $\CN=1$ NEW supergravity.  It would be also interesting to see  the UV behavior of NEW gravity and $\CN=1$ NEW supergravity.

\section*{Acknowledgments}
SH is supported in part by the National Research Foundation of Korea(NRF)  grant funded by the Korea government(MEST) with  the grant number  2009-0074518 and the grant number 2009-0085995. SH, WJ and SHY are supported by the grant number 2005-0049409 through the Center for Quantum Spacetime(CQUeST) of Sogang University. JJ is supported by the National Research Foundation of Korea(NRF)  grant funded by the Korea government(MEST) with  the grant number 2009-0072755.

\appendix
\section{Conventions and Some Useful Formulae}
In this appendix we present our conventions and also give some formulae. 
The $4\times 4$ $\g$-matrices in flat spacetime are defined by the
Clifford algebra,
\beq
\{ \g_{a} \,, \g_{b} \} = 2 \eta_{ab}~,\qquad \eta_{ab}={\rm diag}(-\, +\,+\,+)\,.
\eeq
The $\g$-matrices in curved spacetime are defined by $\g_\m=e^a_\m \g_a$ with $g_{\m\n}=\eta_{ab}e^a_\m e^b_\n $.
Multi-indexed $\g$-matrices are defined by
\bea
\g_{\m\n} &=& \g_{[\m} \g_{\n]} \equiv \half ( \g_{\m} \g_{\n} - \g_{\n} \g_{\m} )\,,\qquad
\g_{\m\n\r} = \g_{[\m}\g_{\n}\g_{\r]} \,,\\
\g_{5} &=&  i \g_0\g_1\g_2\g_3 = i \g_{0123} \,.\nn
\eea
%
The Majorana spinor is defined by
\beq
\bpsi \equiv \psi^{\dagger} i \g^0 = \psi^{T} C\,.
\eeq
Hermitian conjugates of $\g$-matrices are
\beq
\g_{\m}^{\dag} = -\g^{0} \g_{\m} ( \g^{0})^{-1}\,,\qquad
\g_{\m\n}^{\dag} = - \g^{0}  \g_{\m\n} (\g^{0})^{-1}\,,\qquad
\g_{\m\n\r}^{\dag} =  \g^{0}  \g_{\m\n\r} (\g^{0})^{-1}\,,
\eeq
and trasnposes of $\g$-matrices are
\bea
\g_{\m}^T &=& - C \g_{\m} C^{-1} \,, \qquad C^T = - C\,, \\
(C \g^{\m})^T &=& C \g^{\m}\,, \qquad (C \g^{\m\n} )^T = C \g^{\m\n} \,,\nn\\
(C \g^{\m\n\r} )^T &=& - C \g^{\m\n\r}\,,\qquad (C \g^{\m\n\r\sigma} )^T = - C \g^{\m\n\r\sigma}\,.\nn
\eea
Some useful bilinear identities are
\bea
&& \bpsi \lambda = \blambda \psi\,,\qquad \bpsi \g^{\m} \lambda = - \blambda \g^{\m} \psi\,, \qquad \bpsi \g^{\m\n} \lambda = -\blambda \g^{\m\n} \psi \,,\nn\\
&& \bpsi \g^{\m\n\r} \lambda = \blambda \g^{\m\n\r} \lambda\,,\qquad \bpsi \g^{\m\n\r\s} \lambda = \blambda \g^{\m\n\r\s} \psi\,.
\eea
Supersymmetry variations of vielbein and gravitino fields in our convention are given by
\bea  \delta e^{a}_{\mu} &=& \bar{\epsilon}\gamma^a\psi_{\mu}\,, \nn \\
         \delta \psi_{\mu} &=& -\Big(D_{\mu} + \frac{1}{2 \ell}\gamma_{\mu}\Big)\epsilon+\cdots\,.
\eea
The covariant derivatives are defined by
\bea
D_{\m} \psi_{\n} &=& \p_{\m} \psi_{\n} + \frac{1}{4} \omega_{\m}^{~ab}\g_{ab}\psi_{\n}+ \frac{i}{2} A_{\m} \g_{5} \psi_{\n} \,,\\
D_{\m} \chi_{\n} &=& \p_{\m} \chi_{\n}  + \frac{1}{4} \omega_{\m}^{~ab}\g_{ab}\chi_{\n}- \frac{i}{2} A_{\m} \g_{5} \chi_{\n}\,.\nn
\eea
A useful identity for $\chi_0$ is
\beq
\gamma_{\mu}(D^{[\mu}\psi^{\nu]} -\gamma^{[\mu}\chi^{\nu]}_0)=0\,.
\eeq
Riemann tensor and Ricci tensor are defined by
\bea
R^{\r}_{~\s\m\n} &=& \p_{\m} \G^{\r}_{\n\s} -\p_{\n} \G^{\r}_{\m\s} + \G^{\r}_{\m\l}\G^{\l}_{\n\s} - \G^{\r}_{\n\l}\G^{\l}_{\m\s}\,,\\
R_{\m\n} &= &R^{\r}_{~\m\r\n}\,.\nn
\eea

\end{document}